\let\txt=\textstyle
\def\gapp{ {\txt {{\txt >} \atop {\txt \sim}}} } 
\documentstyle[preprint,revtex]{aps}
\begin{document}
\draft
{}.
\vskip 5cm
{\hskip 4 in {UM AC 92-6}\hfil\break}

{\hskip 4 in {December 1992}\hfil\break}
\begin{title}
Structure in a Loitering Universe
\end{title}
\author{Hume A. Feldman and August E. Evrard}
\vskip 1cm
\begin{instit}
Physics Department
University of Michigan
Ann Arbor, MI 48109
\end{instit}
\vskip 2cm
\begin{abstract}
We study the formation of structure for a universe that undergoes a
recent loitering phase. We compare the nonlinear mass distribution to
that in a standard, matter dominated cosmology. The statistical
aspects of the clustered matter are found to be robust to changes in
the expansion law, an exception being that the peculiar velocities are
lower by a factor of $\sim 3$ in the loitering model. Further, in the
loitering scenario, nonlinear growth of perturbation occurs more
recently ($z\sim 3-5$) than in the matter dominated case.  Differences
in the high redshift appearances of the two models will result but
observable consequences depend critically on the chosen form, onset
and duration of the loitering phase.
\end{abstract}
\vskip 0.5cm
\pacs{98.80.Bp  98.80.Dr}
\narrowtext

\section{Introduction}

The standard Freedman--Robertson--Walker (FRW) universe with a hot big
bang is simple and quite successful in predicting the Hubble
expansion, the abundances of light elements and other phenomena.  The
current standard theories for structure formation, in particular
inflation and cold dark matter (CDM), seem to give a reasonable
explanation as to the formation of large scale structure in the
Universe.

The standard cosmology invokes gravitational instability as the
mechanism for structure formation. Forming structure gravitationally
is very attractive, in numerical simulation structure that is
remarkably similar to the one observed seems to form. Further, the
recent COBE\cite{smoot,write} observations of anisotropy of the microwave
background radiation (CMBR) suggests that the primordial Universe was
indeed slightly inhomogeneous and anisotropic. These tiny
inhomogeneities may have been seeds that grew gravitationally to
become the structure we see today.

As successful as it is, the standard model may be in apparent conflict
with some cosmological observations. Recent findings suggest more
inhomogeneities\cite{lss} on large scales $\gapp 10$Mpc and smaller
peculiar velocities\cite{defw} on small scale $\sim1$Mpc than the
unbiased CDM scenario predicts. It might be possible to reconcile some
of the observations with the model, {\it e.g.,} a biased CDM model
gives a better agreement with small scale streaming velocities, but it
makes the large scale power discrepancy worse. Another example is a
flat cosmology with a nonvanishing cosmological constant which helps
ameliorate the large scale velocities\cite{ems}, however it conflicts
with lensing statistics\cite{turner}. Thus it is not clear that CDM
can be reconciled with observations of large scale structure and the
cosmic microwave background radiation temperature variations.

Although gravitational instability has been tested for structure
formation given standard cosmological background, {\it i.e.,} FRW with
CDM\cite{defw}, only a few studies of other cosmologies have been done
\cite{martela,lahav}.  In this paper, we present results
for structure formation in a universe obeying a non--standard
expansion law\cite{loiter}.  The motivation is to test the `stability'
of gravitational instability to changes, not in the initial
conditions, but in the expansion law of the background metric.  A
universe which undergoes a period of `loitering' is an attractive
alternative to standard cosmologies\cite{loiter}.  The loitering
universe scenario is an expanding Friedmann cosmology that undergoes a
fairly recent ($z\sim3-5$) phase of slow expansion.  It is during this
semi--static phase that large--scale structure is formed.  The model
is useful in providing explanations both for the small amplitude of
CMBR fluctuations and for the so--called `age problem' of globular
clusters. Although a period of loitering can be achieved in a
cosmological constant dominated universes\cite{martela}, Durrer \&
Kovner\cite{durrer} have noted that there are serious problems with these
models.

Some researchers\cite{jd} claim that the ages of the oldest globular
clusters are $1.6\times10^{10}$ years, whereas the estimation of the
Hubble constant today varies between ${\rm H}_o^{-1}=1-2\times10^{10}$
years. Since, in a matter dominated universe, the age of the universe
$t_o\leq {\rm H}_o^{-1}$, and in particular, $t_o=2/3{\rm H}_o^{-1}$
for flat ($\Omega=1$) cosmologies, [younger for an open ($\Omega_o<1$)
universe]. The ages of globular clusters pose a real problem,
especially if ${\rm H}_o$ is large (small ${\rm H}_o^{-1}$) as some
recent observations suggest\cite{age}.  The loitering scenario allows
for a universe where $t_o>{\rm H}_o^{-1}$ as can be seen from the
expansion law in figure 1 below.

Until the recent apparent detection of anisotropy in the
CMBR\cite{smoot,write}, it was essential to investigate theories
capable of producing structure from very low amplitude initial
perturbations.Although the COBE result seems to confirm the standard
model, it remains the only detection of the microwave anisotropy.
Until confirmation of this detection and further detections on
different angular scales, alternatives to the standard model should be
considered viable.  During the loitering phase, the density contrast
of initial perturbations grows semi--exponentially with expansion
factor, much faster than the power law dependence in standard FRW
cosmologies.  The rapid growth in amplitude allows non--linear
structures to form at present without producing large Sacks--Wolfe
anisotropies\cite{loiter}.  However, detailed properties of clustering
in the non--linear regime have not yet been investigated for this
model.

In this paper we employ N--body simulations in both loitering and
matter dominated (MD) cosmologies to test statistical properties of the
large--scale matter distribution in the non--linear regime.
We find that the loitering scenario produces structure
remarkably similar to that formed in the MD scenario, with some
interesting differences. For the loitering scenario the Sacks--Wolfe
$\Delta T/T$ is smaller, small scale streaming velocities agree better with
observations, and the universe is old enough to account for the age of the
oldest globular clusters.

The paper is organized as follows: In section II we review the
loitering universe model. In section III, we present details of the
initial conditions and results of the simulations.  We conclude in
section IV.

\section{The Loitering Model}

For the purpose of testing the large scale structure predictions of
the loitering model, we developed a simple model that has all the
interesting features of the loitering scenario\cite{loiter} without
any of the more complicated dynamics involved in a more sophisticated
theory.  Consider a closed universe where, in addition to the
conventional matter ({\it i.e.,} matter with both $p$ and $\rho\ge0$),
there exists a source term having an equation of state $p_s =
\gamma\rho_s$ where $-1\le\gamma<-\frac13$. The energy density of
such matter scales as $\rho_s\propto{1\over a^n}$ where $0 \leq n
< 2$, [$n=3(1+\gamma)$].

If we assume that the additional source term does not interact with
ordinary matter, the only thing that will affect the formation of
structure is the different expansion law, which is why we can use the
simplistic theory as a good approximation for structure formation
studies. This effect is the one we study here. The Einstein equations
for a Friedman cosmology with an additional source term are
\begin{equation}
\left({\dot a\over a}\right)^2 =
	{8\pi\over3}\left(\rho_{_{\rm T}}
	- {k\over a^2}\right)
 	= {8\pi\over3}\left({\rho_o\over a^3}
	+ {\rho_s\over a^n}-{k\over a^2}\right)
\label{ee1}
\end{equation}
and
\begin{equation}
\frac{\ddot a}{a}=
  \frac{4\pi G}{3}(\rho_{_{\rm T}}+3p)\
	=
	\frac{4\pi G}{3}\left(
		{\rho_o\over a^3} +
		{(n-2) \rho_s\over a^n}
	\right) \ ,
\label{ee2}
\end{equation}
where $\rho_o$ and $\rho_s$ are constants and $k=1$ (closed universe).
There is an implicit assumption that at late times the conventional
matter content of the universe is in the form of cold matter, {\it
i.e.,} $\rho_{\rm cm}\propto\,a^{-3}$. When all the terms in the right
hand side of Eq.  (\ref{ee1}) are of comparable magnitudes, a period
of coasting may result during which $\dot a\simeq {\rm constant}\ll1$
and $\ddot a
\simeq 0$. After the `loitering' phase the universe goes into a late,
power law `inflationary' phase.

The expansion law of the model we consider here and the standard FRW
cosmology expansion law are shown in Figure \ref{expansion}. For the
run described in the paper we chose the parameters to be $n=1.2$ and
$\rho_s/\rho_o=0.32$. This choice of parameters implies that the age
of the universe is $t_o\simeq8/3{\rm H}_o^{-1}$ whereas in a matter
dominated universe $t_o=2/3{\rm H}_o^{-1}$, (see Figure \ref{age})
this solves the `age problem'.  Clearly the form of the expansion law
depends on the choice of these parameters. Ideally, the values of
these parameters should be determined from the nature of the matter
described by the source term.  The parameters we chose lead to a
rather short and recent period of loitering. The onset, duration and
form of the expansion law can be manipulated by choosing different
values of $n$ and $\rho_s$.

We assume that the matter which undergoes gravitational clustering has
negligible pressure so that the gravitational instability is not
hindered by sound waves or free--streaming. We parameterize the amount
of the clustering matter (CM) by
\begin{equation}
\Omega_{\rm cm} =
	{\rho_{\rm cm}\over\rho_{\rm crit}}\ ; \qquad
\rho_{\rm crit}\equiv{3\over8\pi G}\left({\dot a\over a}\right)^2\ ,
\label{omega}
\end{equation}
where $\rho_{\rm cm}$ is the density of CM. We use as our working
hypothesis\cite{LB86} that $\Omega_{\rm cm}\ge0.1$. Since, during the
late inflationary stage, $\Omega_{\rm cm}$ starts to drop, we define
today by the requirement that the ratio between the amount of matter
that clusters to the critical density is not be smaller than that. Our
constraint on $\Omega_{\rm cm}$ means that today cannot be too late in
this model.

As the expansion of the universe slows down, the growth of the
density contrast: $\delta = \delta\rho_{\rm cm}/\rho_{\rm cm}$ speeds up. This
can be readily seen from the Bonner equation describing the growth of the
density
contrast in an expanding universe:
\begin{equation}
\ddot\delta+2\frac{\dot a}{a}\dot\delta=
 4\pi G\rho_{\rm cm}\delta\ ,
\label{bonner1}
\end{equation}
a change in variables $\delta = {\tilde\delta/a}$ transforms this
equation to
\begin{equation}
\ddot{\tilde\delta} - (4\pi G\rho_{\rm cm} +{\ddot a\over a})\tilde\delta = 0\
{}.
\label{bonner2}
\end{equation}
If both ${\dot a\over a}$ and ${\ddot a\over a}$ are small compared to
$4\pi G\rho_{\rm cm}$, Eq. (\ref{bonner2}) can be solved in the WKB
approximation, giving
\begin{equation}
\delta = {1\over a}\exp\left(\int\sqrt{4\pi G\rho_{\rm cm}}\,dt\right) =
     {1\over a}\exp\left(\int\sqrt{{3\over2}\Omega_{\rm cm}}\,d\ln a\right)\ .
\label{delta}
\end{equation}
Thus, as long as $\Omega_{\rm cm}\gg1$ ($\dot a/a \ll\rho_{\rm cm}$), the
perturbations grow rapidly.  This represents an enormous speed up over
the growth rate of perturbations in a matter dominated FRW universe,
in which $\delta \propto a$.

\section{Simulation Results}

We use N--body simulations to investigate details of the non--linear
mass distribution.  Initial conditions were constructed by random
sampling a scale--free power spectrum $P(k) d^3k = A k^n$ with spectral
index $n = -1$ and amplitude $A$ discussed below.
The spectral index chosen $n = -1$ is approximately the slope of the
CDM spectrum on the scale of rich clusters of galaxies $\sim 10^{15}
{\rm M}_\odot$.  Sampling is done
at $64^3$ discrete modes appropriate to the Fourier domain of a
periodic, $64^3$ spatial grid.  A  displacement field is generated on
this grid from the spectral information via use of the Zel'dovich
approximation\cite{edfw}.  The initial particle positions and momenta
are fed as initial conditions into a P3M N--body code\cite{ee}.
The code calculates gravitational forces using Fourier transforms on
large scales and pairwise summations on small scales.
Pairwise forces are softened on small scales using a Plummer potential
with softening parameter $\varepsilon = 0.00125 L$
where $L$ is the length of the periodic cube.  Energy was
conserved to better than $1\%$ in the Layzer--Irvine
equation\cite{edfw} for both of the runs discussed below.

Two simulations were performed using initial conditions differing only
in their {\it rms} displacement field amplitudes.  During the
loitering phase, perturbations will grow at a greatly enhanced rate
relative to the MD case. Since we did not know {\it a priori} the
endpoint of the loitering run, we arbitrarily scaled down the initial
amplitude of the displacement field by a factor 16 for the loitering
run relative to the MD run.  Time was used as the expansion variable
with 1500 time steps used to evolve the systems to their endpoints.
The endpoint for the MD run was chosen when the variance of mass
fluctuations within a sphere of radius $L/8$ was unity.
The endpoint of the loitering model
was chosen as the epoch in which the mass correlation function matched
that of the MD run.  This process involved some trial and error using
$32^3$ particle runs.

Figure \ref{slice} shows projections of a slice of dimension $L/2
\times L/2\times L/5$ for the two runs at the final time and at an
epoch corresponding to a redshift $z = 4$.  Note the suppression of
clustering in the loitering model at $z = 4$.  As can be seen from
Figure \ref{expansion}, this epoch is close to the beginning of the
loitering phase when dramatic perturbation growth sets in.  By the
final epoch, structure in the two models is nearly identical.  Figure
\ref{mv} displays the mass correlation function and pairwise velocity
statistic in dimensionless units. Remarkably, the correlation
functions of the loitering and MD runs are almost identical.  The
small difference at separations $r/L < 0.002$ are not resolved by
these experiments.  Comparison between the $32^3$ and $64^3$ loitering
runs indicates that convergence is achieved beyond about two softening
lengths.

A big difference, however, exists in the pairwise velocity statistic.
The amplitude of velocities in the loitering run is suppressed by a
factor $\sim 3$ relative to the MD run.  It is well known that $\Omega
= 1$ MD models have a `velocity problem' on small scales\cite{defw}.
The problem can be expressed very simply along the following lines.
If galaxies trace mass so that a critical density
of dark matter is in halos around bright galaxies
then the expected velocity on a spatial scale
$R$ can be inferred from application of the
virial theorem
\begin{equation}
v \simeq \sqrt{3\Omega_{\rm cm} \over 8 \pi n_* R^3} {\rm H}_o R
\label{virial}
\end{equation}
where $n_* \simeq 0.002 h^3$ Mpc$^{-3}$ is the observed number density
of bright galaxies, $h = {\rm H}_o/(100 {\rm km/s/Mpc})$.  This leads
to the result that one expects velocity amplitudes of $\sim 1000 {\rm
km/s}$ on scales of $1 {\rm Mpc}$.  This rough calculation is
supported by numerical experiments\cite{ee,ae} which invoke biased
galaxy formation as a partial solution.  Even biased models do not
bring the velocities down to the observed level of $\sim 300 {\rm
km/s}$.  The loitering model has a low density of clustered matter and
therefore behaves like a low $\Omega$ universe from the point of view
of velocities. Linear perturbation analysis\cite{martelb} confirms
that the amplitude of the velocity field depends almost entirely on
the value of $\Omega_{\rm cm}$ and only weakly on the value of the
source term which acts as an effective cosmological constant at every
time slice.  Although attractive on small scales, this feature of the
model may be a liability on $10 {\rm Mpc}$ scales where velocity
fields of $\gapp 500 {\rm km/s}$ are claimed to exist\cite{dressler}.

In Figure \ref{group}, we plot the group multiplicity function for the two
models at two different redshifts. The groups have been defined using
a traditional friends--of--friends algorithm which joins common members
of pairs with separations $r < \eta N^{-1/3}$.  We use $\eta = 0.15$
to pick out groups at a density contrast of several hundred.  Although
there is a large difference between the cumulative numbers of
collapsed objects at $z>3$, the numbers are similar at $z=0$.
However, the loitering model has a slightly flatter low mass slope and
has somewhat more high mass systems at the present.

\section{Summary and Discussion}

We have investigated the formation of structure for a universe that
undergoes a recent loitering phase and compared it to a conventional,
matter dominated universe.  The universe under investigation is much
older than the matter dominated one, $t_o\simeq8/3{\rm H}_o^{-1}$.
Further, we find that despite the different rates of perturbation
growth in the linear regime, the non--linear structure in the two
models is remakably similar.  In particular, the shape of correlation
function of the clustered matter and the mass functions of collapsed
objects are nearly identical at the present epoch.  However, because
the loitering universe contains a factor ten less clustered matter
than the flat, MD model, the peculiar velocities are reduced by a
factor of $\sqrt{10}$.

Observed peculiar velocities on small scales ($1$Mpc) favour models
where the density of clustered matter is low. However, observed
velocities on large scales ($\gapp 10$Mpc) favour high density models.  The
loitering model fails to generate high enough velocities on large
scales but provides good agreement with observations on small scales.
The flat MD model has the inverse problem of generating high amplitude
velocities on all scales.  At present, there are no models capable of
reproducing the observed velocities on all scales.

At high redshift, differences in the mass distributions between the loitering
and MD models become apparent.  For example, Figure \ref{group} shows an order
of magnitude difference in the expected abundance of massive, collapsed
objects.
The abundance of quasars at high redshift may, in principle, provide a
distinguishing signature between the models.  However, detailed predictions
from
the loitering universe will depend critically on the choice of
parameters.

In a broader context, this study addresses the question of the
`stability' of gravitational instability as a mechanism for structure
formation under changes in the assumed expansion law of the universe.
We have found the resultant non--linear structure to be remarkably
robust.  To distinguish between models invoking different expansion
laws, observations of structure at both low and high redshifts is
required.

\vskip 1cm
\noindent{\bf Acknowledgement:}  H.A.F. was supported in part by
National Science Foundation grant NSF--92--96020. A.E.E. was supported
in part by NASA Theory grant NAGW--2367.

\vfill\eject

\figure{The expansion laws for the two simulations.\label{expansion}}
\figure{The age of the universe for the two simulations in units of
$2/3{\rm H}^{-1}$.\label{age}}
\figure{Projections of a slice of dimension $L/2 \times L/2
\times L/5$ for the loitering run (left) and the MD run (right) at the
final time (bottom) and at an epoch corresponding to a redshift $z =
4$.  Note the suppression of clustering in the loitering model
at $z = 4$ (top).\label{slice}}
\figure{Displays the mass correlation function (top) and pairwise velocity
statistic (bottom) in dimensionless units. The correlation functions
of the loitering and MD runs are almost identical.  The small
difference at separations $r/L < 0.002$ are not resolved by these
experiments as comparison between the $32^3$ and $64^3$ loitering runs
indicates. The pairwise velocity statistics show that the amplitude of
velocities in the loitering run is suppressed
relative to the MD run.\label{mv}}
\figure{The group multiplicity function for the two
models at two different redshifts. We pick out groups at a density
contrast of several hundred.  Although there is a large difference
between the cumulative numbers of collapsed objects at $z>3$, the
numbers are similar at $z=0$.\label{group}}

\end{document}